\documentclass[english]{article}
\usepackage[T1]{fontenc}
\usepackage[latin9]{inputenc}
\usepackage{float}
\usepackage{graphicx}
\usepackage{esint}
\usepackage{subscript}

\makeatletter

\providecommand{\tabularnewline}{\\}

\newcommand{\lyxaddress}[1]{
	\par {\raggedright #1
	\vspace{1.4em}
	\noindent\par}
}

\makeatother

\usepackage{babel}
\begin{document}
\title{Camel back shaped Kirkwood-Buff Integrals}
\author{Aurélien Perera$^{\dagger}$, Martina Požar$^{\ddagger}$ and Bernarda
Lovrin\v{c}evi\'{c}$^{\ddagger}$}
\maketitle

\lyxaddress{$^{\dagger}$Laboratoire de Physique Théorique de la Matière Condensée
(UMR CNRS 7600), Sorbonne Université, 4 Place Jussieu, F75252, Paris
cedex 05, France.}

\lyxaddress{$^{\ddagger}$Department of Physics, Faculty of Science, University
of Split, Ru\dj era Boškovi\'{c}a 33, 21000, Split, Croatia.}
\begin{abstract}
Some binary mixtures, such as specific alcohol-alkane mixtures, or
even water-tbutanol, exhibit two humps \textquotedblleft camel back\textquotedblright{}
shaped KBI. This is in sharp contrast with usual KBI of binary mixtures
having a single extremum. This extremum is interpreted as the region
of maximum concentration fluctuations, and usually occurs in binary
mixtures presenting appreciable micro-segregation, and corresponds
to where the mixture exhibit a percolation of the two species domains.
In this paper, it is shown that two extrema occur in binary mixtures
when one species forms \textquotedbl meta-particle\textquotedbl{}
aggregates, the latter which act as a meta-species, and have their
own concentration fluctuations, hence their own KBI extremum. This
\textquotedbl meta-extremum\textquotedbl{} occurs at low concentration
of the aggregate-forming species (such as alcohol in alkane), and
is independant of the other usual extremum observed at mid volume
fraction occupancy. These systems are a good illustration of the concept
of the duality between concentration fluctuations and micro-segregation. 
\end{abstract}

\section{Introduction}

The so-called Kirkwood-Buff integrals (KBI)\cite{KBI_Kirkwood_Buff_seminal,Textbook_Ben_Naim_inverse_KBI}
are defined as the integrals of the species-species pair correlation
functions$G_{ab}=\frac{1}{\Omega}\int d\mathbf{X}_{1}d\mathbf{X}_{2}\left[g_{ab}(\mathbf{X}_{1},\mathbf{X}_{2})-1\right]$
where $\mathbf{X}_{i}$ is the set of position, and if required, the
orientations, of molecule $i$, where $a$ and $b$ designates the
species indexes ($\Omega=V\omega^{2}$, where $\omega$ is the angular
integral, equal to $4\pi$ or $8\pi^{\ensuremath{2}}$, depending
on the symmetry of the molecules). It can be shown \cite{Textbook_Hansen_McDonald,2019_ethMeth_KBI}
that this integral is identical to that involving the pair correlation
between any two atoms belonging to each molecules. Following this,
the KBI are more simply defined as 
\begin{equation}
G_{ab}=4\pi\int_{0}^{\infty}drr^{2}\left[g_{i_{a}j_{b}}(r)-1\right]\label{KBI}
\end{equation}
where $i_{a}$and $j_{b}$ designate any two atoms $i$ and $j$ of
, respectively, species $a$ and $b$, and $g_{i_{a}j_{b}}(r)$ the
atom-atom pair correlation function.The Kirkwood-Buff theory relates
specific thermodynamic properties, such as the compressibility, or
the partial molar volumes, for example, to be related to these integrals,
thus providing an appealing link between macroscopic measurable properties
to the microscopic structure, albeit in an integrated form, where
all microscopic details are lost. Ben-Naim \cite{KBI_Ben_Naim_Inversion_water_eth},
Matteoli and Lepori \cite{KBI_Matteoli_Lepori} and other authors
\cite{KBI_Marcus_1991,KBI_Marcus_1991_FT,KBI_inversion_Pandey,KBI_Smith_inversion,KBI_Shulgin_water_alcohols}
have managed to invert these relations, providing a way to calculate
these integrals from the experimentally available data on compressibility,
partial molar volume and vapour pressure or Gibbs free-energy. The
corresponding expressions for a binary mixture ($a,b$) are well known
\cite{AUP_alcohol_KBI}: 
\begin{equation}
G_{aa}=G_{ab}+\frac{1}{x_{b}}\left[\frac{\bar{V}_{b}}{D}-V\right]\label{Gaa}
\end{equation}
\begin{equation}
G_{ab}=\frac{\chi_{T}}{k_{B}T}-\frac{\bar{V}_{a}\bar{V}_{b}}{VD}\label{Gab}
\end{equation}
were $x_{b}$ is the mole fraction of component $b$, $\bar{V}_{c}$
is the partial molar volume of component $c$ ($c=a$ or $b$), $V$
is the molar volume, $\chi_{T}$is the isothermal compressibility
(with $T$ the temperature and $k_{\ensuremath{B}}$ the Boltzmann
constant) and $D$ is related to the concentration fluctuations and
given by 
\begin{equation}
D=x_{i}\left(\frac{\partial\beta\mu_{i}}{\partial x_{i}}\right)_{TP}\label{D}
\end{equation}
where $\mu_{i}$ is the chemical component of species $i$ and $\beta=1/k_{B}T$
is the Boltzmann factor.

Fig.1 of the seminal Matteoli and Lepori paper \cite{KBI_Matteoli_Lepori}
displays KBI for various types of aqueous mixtures, and it can be
seen that most of them have a single extremum, which is positive for
like KBI $G_{aa}$ and negative for then unlike $G_{ab}$ with $(a\neq b)$.
This extremum has its origin in the shape of the coefficient $D$
above, which is often a U-like shaped curve with a single minimum.
An example of such typical KBI and $D$ are given in Fig.1, and other
similar examples can be found in various place in the literature such
as \cite{KBI_Donkersloot_WatEth_WatMeth,KBI_Nishikawa_EthWat_Xray,KBI_Marcus_2001,KBI_Marcus_PCCP,AUP_alcohol_KBI}.
Note that, while for a given mixture, the $D$ coefficient has a single
extremum, the corresponding extrema in the KBI are not necessarily
at the same x-position, as illustrated in Fig.1, as Eqs.(\ref{Gaa},\ref{Gab})
tend to alter this position from that of D.

\begin{figure}
\begin{centering}
\includegraphics[scale=0.26]{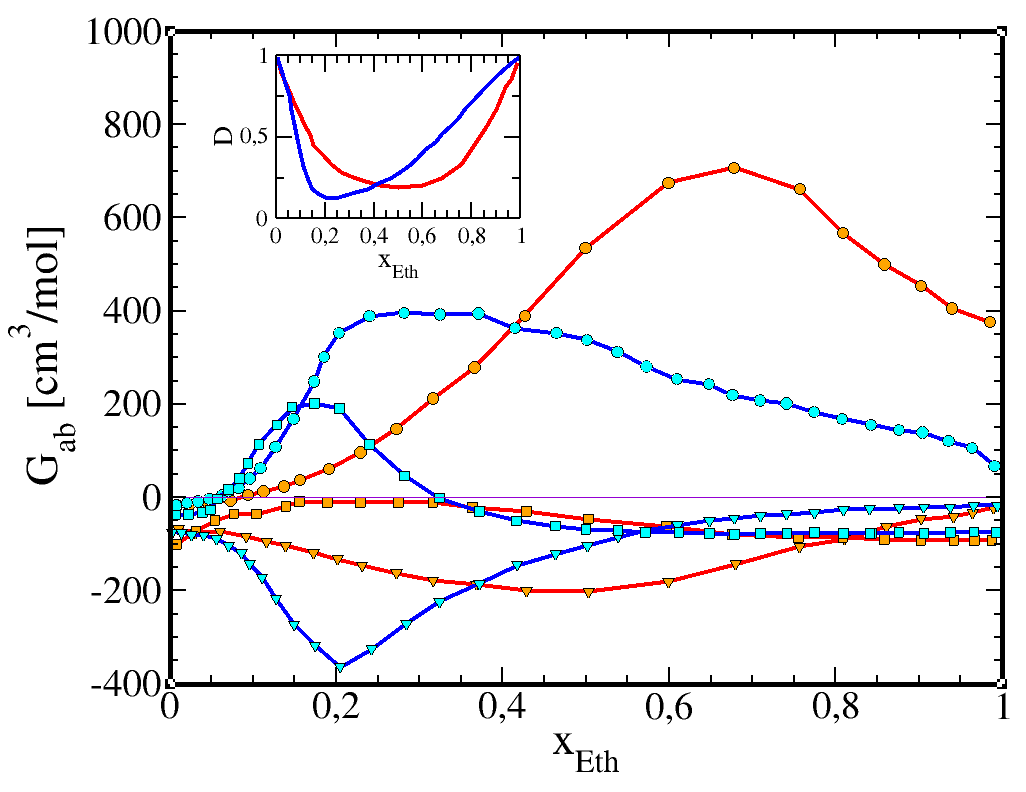}
\par\end{centering}
\caption{Examples of single extremum KBI. Red curves experimental KBI curves
for aqueous-acetone from Ref.\cite{AUP_acetone_Koga}; blue curves
for aqueous-2propanol from Ref.\cite{AUP_alcohol_KBI}. Circles for
Water-water KBI, squares for cross water-solute KBI, and triangles
for solute-solute KBI. The inset shows the corresponding D (Eq.(\ref{D}))
with same color codes, both showing a single minimum.}
\end{figure}

A single extremum for $D$ is expected for a curve which reaches the
same value at both end points. In fact, Eq.(\ref{D}) can be shown
(see Appendix) to be related to the stability limit criteria for a
binary mixture, that is precursor for demixing, namely 
\begin{equation}
D\propto C_{11}C_{22}-C_{12}^{2}\label{D-OZ}
\end{equation}
where $C_{ij}=\delta_{ij}-\sqrt{\rho_{i}\rho_{j}}\tilde{c}_{ij}(k=0)$
, with the $\tilde{c}_{ij}(k=0)$ being the integrals of the direct
correlation functions and i,j refer to species index.Indeed, when
$D\rightarrow0$ the mixture becomes mechanically unstable \cite{Textbook_Hansen_McDonald},
such when approaching a spinodal or a critical point. Therefore, a
single extremum in $D$ can be physically interpreted as the point
where concentration fluctuations are maximal, and with the eventuality
that when these grow beyond the mechanical stability, the mixture
undergoes demixing. Subsequently, the extremum in the KBI is a signature
of maximal concentration fluctuations. Since many binary mixtures,
such as aqueous mixtures, for example, show stable local nano-segregation
of constituents \cite{MH_Soper_Nature,MH_Guo_Methanol_water,MH_Allison_Meth_Water_clusters,AUP_ethWat_JMolLiq,2016_PCCP_MH_Versus_Clust},
the extremum can be interpreted as maximal amplitude in local segregation.

In this context, it came as a surprise that Jan Zielkiewicz published
experimental KBI for various alcohols-heptane mixtures \cite{KBI_Zielkiewicz},
which clearly showed the existence of two extremum. In Fig.2 of his
paper, he shows KBI written as $\rho_{a}G_{ab}$ where $\rho_{a}=N_{a}/V$
is the partial density of species $a$ ($N_{a}$ is the number of
particle in volume $V$), which amplifies the second extremum, the
resulting KBI looking like ``camel back'' shaped, as opposed to
the usual single extremum KBI which are ``dromadary camel back''
shaped.

The purpose of this paper is to understand the origin of this dual
extremum in terms of fluctuations. The driving idea is the following.
It is well known that alcohols generally tend to cluster their hydroxyl
groups into linear patterns, forming chains, loops, etc... \cite{ExpScattNartenEthMeth,ExpScattFinnsMonools,ExpScattJoarderAlcohols,ExpScattMatijaMonools,2020_alcohols_Chris,2021_n_octanols_Chris}
When put in small concentration in an inert alkane solvent, such as
heptane, for example, small ``chain-micelles'' of alcohols form
and act as independant meta-particles. Consequently, they exhibit
their own concentration fluctuations, which corresponds to one of
the observed extrema in the corresponding KBI. With increasing concentration,
the frontier between well separated micelles and alcohol micro-segregated
domains become loose, and the mixture becomes a simple micro-segregated
mixture, which has its own independent maximum in concentration fluctuations.
It is not obvious that such scenario should happen, and the purpose
of this paper is to use computer simulation to provide evidence that
this scenario is correct, principally through the analysis of pair
correlation functions and the KBI.

\section{Theoretical and simulation details}

The study of equilibrium locally micro-segregated mixtures has proven
to be very difficult in the past decades, particularly in the case
of aqueous mixtures. Perhaps a canonical example is that of aqueous
\emph{tert}-butanol (TBA) mixtures, for which the force field induced
slow demixing at very small TBA concentrations around $x_{\mbox{TBA}}\approx0.01$
was shown \cite{MH_Nico_TBA_Wat,MH_Bagchi_TBA_Wat,SIM_Patey_Gupta_TBA,SIM_Patey_Overduin_TBA,AUP_Patey_TBA}
to rule out many classically robust force field models such as OPLS
\cite{FF_OPLS_hydrocarbons,FF_OPLS_alcohols_1} or TraPPE \cite{FF_Trappe_hydrocarbons,FF_Trappe_Alcohols}.
One may even consider this problem to remain unsolved. Similarly,
the water-tetrahydrofuran mixture remains difficult to simulate, in
particular for putting into evidence the loop coexistence phase diagram
\cite{EXP_THF_WAT_LLC,EXP_THFWat_2,SIM_THF_Wat}. One of the principal
obstacle for successfully simulating such mixtures is the fact that
both water and the alcohol tend to form hydrogen bonded clusters,
and it would seem that it is precisely the competition between these
cluster formation which tend to drive the simulated mixture into demixing,
often after unusually very long times.

These problems might be avoided if water is replaced by a more inert
solvent such as an alkane. In a previous work \cite{2015_PCCP_benzMH},
we have studied ethanol-benzene mixtures and found that the description
of the strong micro-heterogeneity (MH) in the low ethanol content
region posed statistical issues which necessitated large size simulations.
We expect to find similar issues in the present study.

Herein, we focus on the ethanol-heptane mixtures, simulated in the
following range of ethanol mole fractions: 0.02, 0.05, 0.1, 0.2, 0.3,
0.4, 0.5, 0.6, 0.7, 0.8 and 0.9. All of the mixtures contain N=16000
particles, in order to properly describe extented MH. The program
package Gromacs was used to perform molecular dynamics simulations
\cite{MD_gromacs_4_5,MD_gromacs_parallelization}. The simulation
protocol has been the same for all mole fractions of ethanol. The
initial random configurations of 16000 molecules were created with
the program Packmol \cite{MD_Packmol}. These initial configurations
were first energy minimized and then equilibrated in the NpT ensemble
5 ns. The length of the production runs varied for each mole fraction
of ethanol and can be found in Table \ref{tab1_production_times}.
The shortest production runs of 10 ns sampled on average 1000 configurations,
whereas the longest runs of 35 ns sampled more than 3500 configurations.

\begin{table}[h]
\begin{tabular}{|c|c||c|c|}
\hline 
Mole fractions x\textsubscript{ETH} & Production run times & Mole fractions x\textsubscript{ETH} & Production run times\tabularnewline
\hline 
\hline 
0.02 & 10 ns & 0.5 & 25 ns\tabularnewline
\hline 
0.05 & 35 ns & 0.6 & 20 ns\tabularnewline
\hline 
0.1 & 35 ns & 0.7 & 15 ns\tabularnewline
\hline 
0.2 & 30 ns & 0.8 & 10 ns\tabularnewline
\hline 
0.3 & 35 ns & 0.9 & 10 ns\tabularnewline
\hline 
0.4 & 20 ns &  & \tabularnewline
\hline 
\end{tabular}

\caption{\label{tab1_production_times}Production run times for the simulated
ethanol-heptane mixtures.}
\end{table}

The TraPPE forcefield for heptane \cite{FF_Trappe_hydrocarbons} and
ethanol \cite{FF_Trappe_Alcohols} was used throughout the range of
ethanol mole fractions. The mixtures were simulated at T = 300 K and
p = 1 bar. Temperature was maintained constant using mostly the v-rescale
\cite{MD_thermo_Vrescale} or Nose--Hoover \cite{MD_thermo_Nose,MD_thermo_Hoover}
thermostat, while pressure was maintained with the Parrinello--Rahman
barostat \cite{MD_barostat_Parrinello_Rahman_1,MD_barostat_Parrinello_Rahman_2}.
The temperature algorithms had a time constant of 0.2 ps, while the
pressure algorithm was set at 2 ps. The integration algorithm leap-frog
\cite{MD_INT_Leapfrog} was used at every time-step of 2 fs. The short-range
interactions were calculated within the 1.5 nm cut-off radius. The
electrostatics were handled with the PME method \cite{MD_PME}, and
the constraints with the LINCS algorithm \cite{MD_constraint_LINCS}.

The snapshots were made with VMD \cite{VMD_1}.

\section{Results}

\subsection{Experiments}

The experimental KBI data for the ethanol-heptane mixtures scanned
from Zielkiewicz paper are reported as lines in the main panel of
Fig.2.

\begin{figure}[H]
\begin{centering}
\includegraphics[scale=0.26]{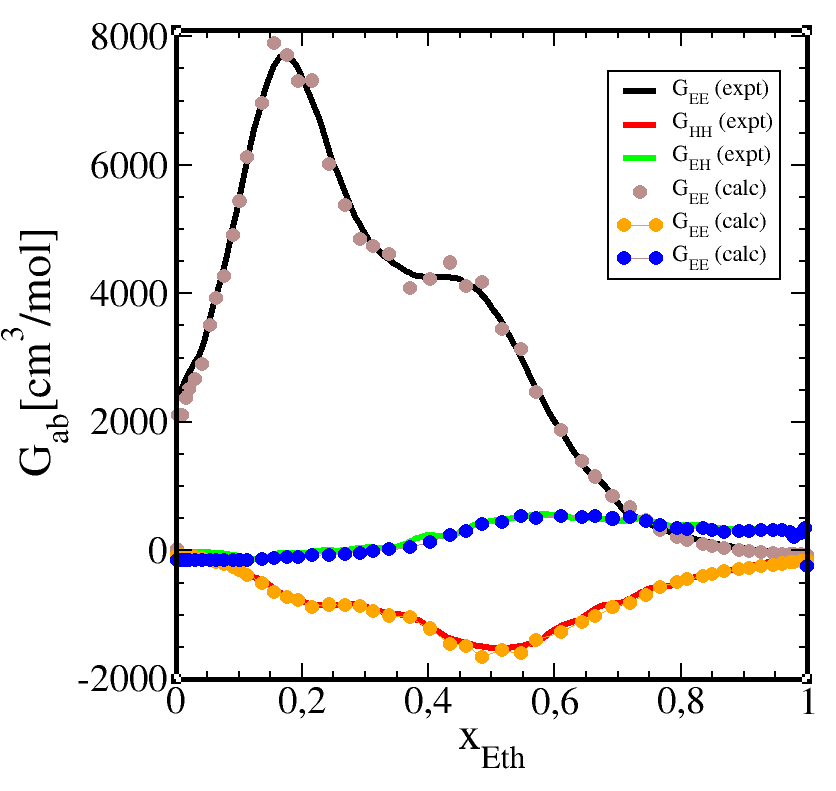}
\par\end{centering}
\caption{Experimental ``camel back shaped'' KBI from Ref.\cite{KBI_Zielkiewicz}
for the ethanol-heptane mixtures, reproduced as lines, with symbols
for recalculated values (see text) for consistency check.}
\end{figure}

In order to make sure that these results are fully consistent with
one another, we have inverted Eqs(\ref{Gaa},\ref{Gab}) in order
to extract $D$, using the following assumptions. The partial molar
volumes have been replaced by the molar volume of each species, thus
making these quantities independent of the concentrations. The isothermal
compressibility term has been neglected (assuming the incompressibility
of the liquid mixtures). Finally, the excess volume of the mixture
has been set to be a linear function of the pure liquid volumes, hence
neglecting the excess volumes. These are usually found to be less
that 0.3cm\textasciicircum 3/mol. All 3 assumptions can be justified
only by the resulting KBI (as was proven by us for aqueous alcohol
mixtures in Ref. \cite{AUP_alcohol_KBI,AUP_ethWat_JMolLiq}). From
Eqs(\ref{Gaa},\ref{Gab}) we have extracted the following equivalent
expressions for $D$ 
\begin{equation}
D_{ab}=-\frac{V_{a}B_{b}}{VG_{ab}}\label{Dab}
\end{equation}
\begin{equation}
D_{aa}=\frac{x_{b}V_{b}-x_{a}V_{a}}{f_{1}-f_{2}}\label{Daa}
\end{equation}
with $f_{1}=x_{a}x_{b}\left(G_{aa}-G_{bb}\right)$ and $f_{2}=V(x_{a}-x_{b})$.
The resulting values are reported in Fig.3.

\begin{figure}[H]
\begin{centering}
\includegraphics[scale=0.26]{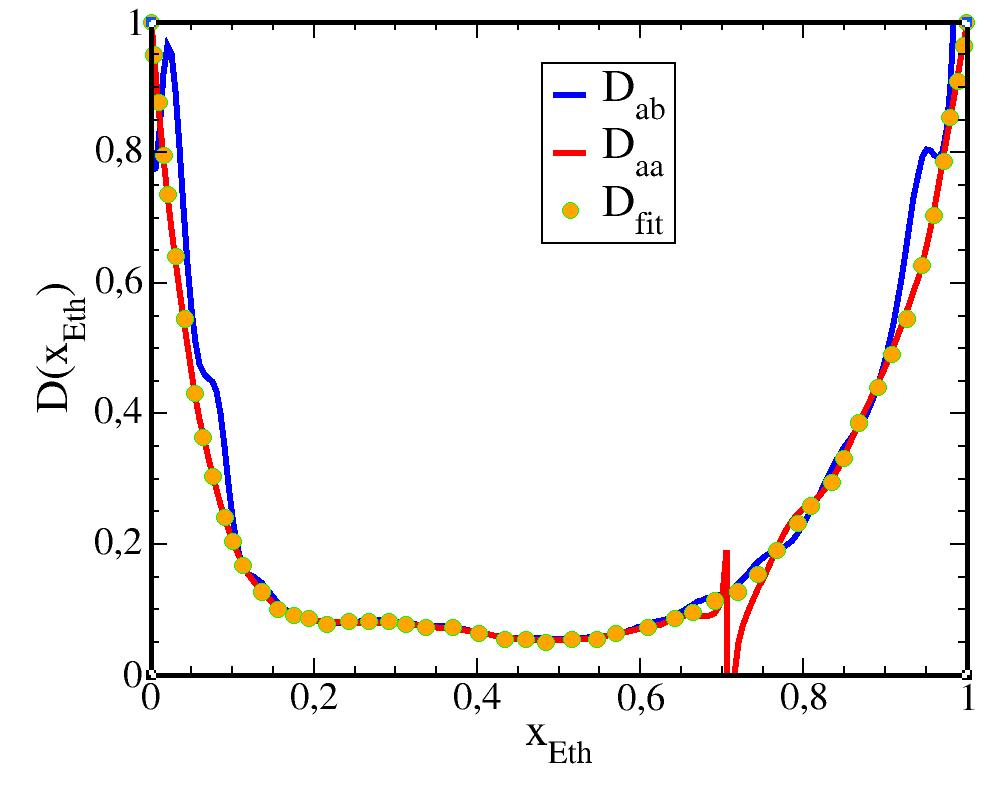}
\par\end{centering}
\caption{Dual minima D for the ethanol-heptane mixtures, as calculated from
the experimental results of Ref.\cite{KBI_Zielkiewicz} (see Eqs.(\ref{Dab},\ref{Daa})
in the text).}
\end{figure}

It is seen that they are quite similar, with the denominator in Eq.(\ref{Daa})
producing a singularity at $x_{\mbox{Eth}}\approx0.7$. The dots represent
the values which have been selected to best represent a good compromise
between the two sets. The most prominent feature is the very apparent
existence of the 2 extrema, one at $x_{\mbox{Eth}}\approx0.2$ and
the other at $x_{\mbox{Eth}}\approx0.5$. Interestingly, both evaluations
of $D$ coincide almost perfectly in the entire range of the minimum
part of $D$. Using these values of the extracted D function, we return
to evaluate back the 3 KBI integrals, by still preserving the 3 approximations
mentioned above. The resulting values are reported in Fig.2 as dots,
and are seen to nicely superpose to the original data. This simple
exercise proves that a single form of $D$, which must be quite close
to the one reported in the inset, must have served in Ref.\cite{KBI_Zielkiewicz}
to calculate the KBI, and moreover, that the use of true partial molar
volumes, compressibilities and excess volume did not affect much the
final shapes. We also prove that the 2 extrema of the KBI originate
from the 2 extrema observed in $D$.

\subsection{Computer simulations}

\subsubsection{Snapshots}

When mixtures present strong local heterogeneity, it is often instructive
to look at snapshots in order to visually appreciate the nature and
the extent of the concentration dependence of the spatial segregation.
Fig.4 shows snapshots of 4 typical concentrations of ethanol, with
different styles highlighting the morphological changes in the ethanol
clustering in heptane solvent.

\begin{figure}[H]
\begin{centering}
\includegraphics[scale=0.3]{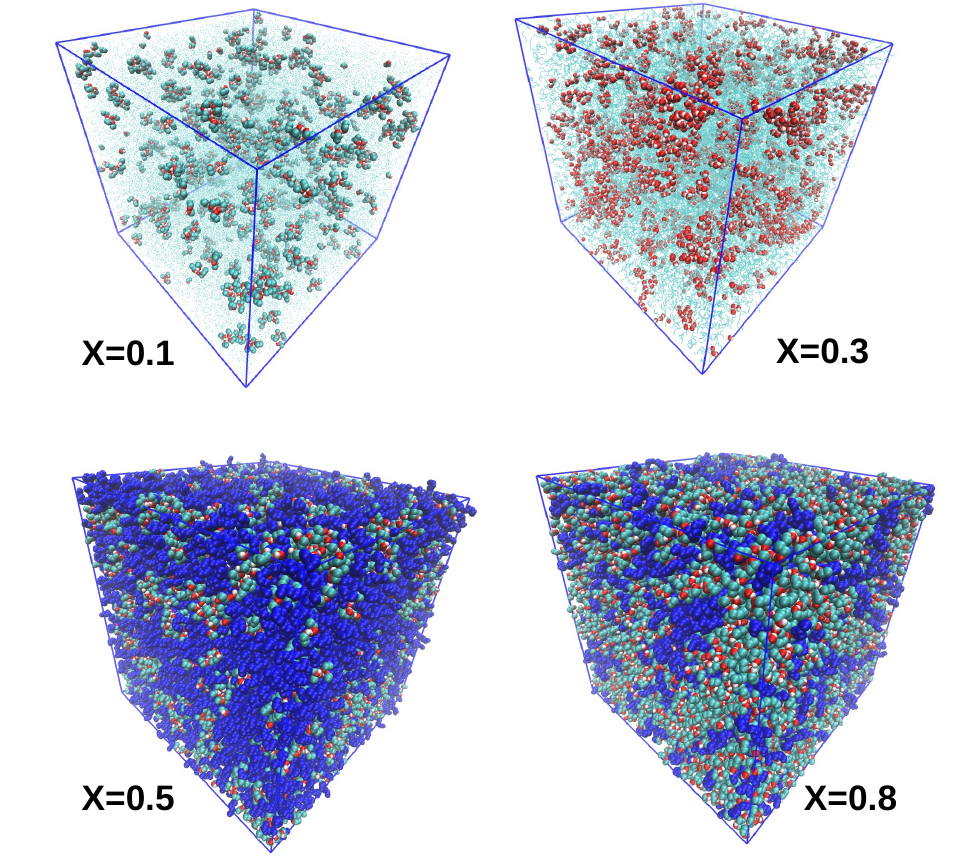}
\par\end{centering}
\caption{Snapshots of the ethanol-heptane mixtures for 4 typical ethanol concentrations.
Ethanol oxygen, hydrogen and carbon groups, in red, white and cyan,
respectively, and heptane carbon groups in blue. Heptane is shown
as ghost pale blue for $x=0.1$ and $0.2$. Ethanol carbon groups
are also shown as ghost pale blue for $x=0.$2. (see text for details).}
\end{figure}

For ethanol concentration $x=0.1$, full ethanol molecules are shown
(oxygen atom in red, hydrogen in white and carbon groups in cyan)
while heptane molecules are shown in transparent mode. It can be seen
that small droplets of ethanol float in the midst of heptane solvent.
For $x=0.$3, only the hydroxyl groups are fully shown, with all carbon
groups of both species shown in transparent mode. This way, hydroxyl
clusters and their distribution are highlighted. For $x=0.5$ and
$x=0.$8, ethanol molecules are fully shown like for $x=0.1$, and
heptane molecules are shown in blue. These last 2 snapshots highlight
the clear micro-segregation of both species.

Fig.5 shows a zoom on the hydroxyl group clustering for the same 4
ethanol concentrations highligted in Fig.4. For $x=0.1$ one sees
that the small ethanol droplets of Fig.4 consist in fact in pentameric
rings of hydroxyl groups, forming ethanol pentameric ``micelles''.
This structure vanishes for the other higher ethanol concentrations
shown in Fig.5, replaced by more complex cluster structures, such
as globules and various type of chains conformations..

\begin{figure}
\begin{centering}
\includegraphics[scale=0.3]{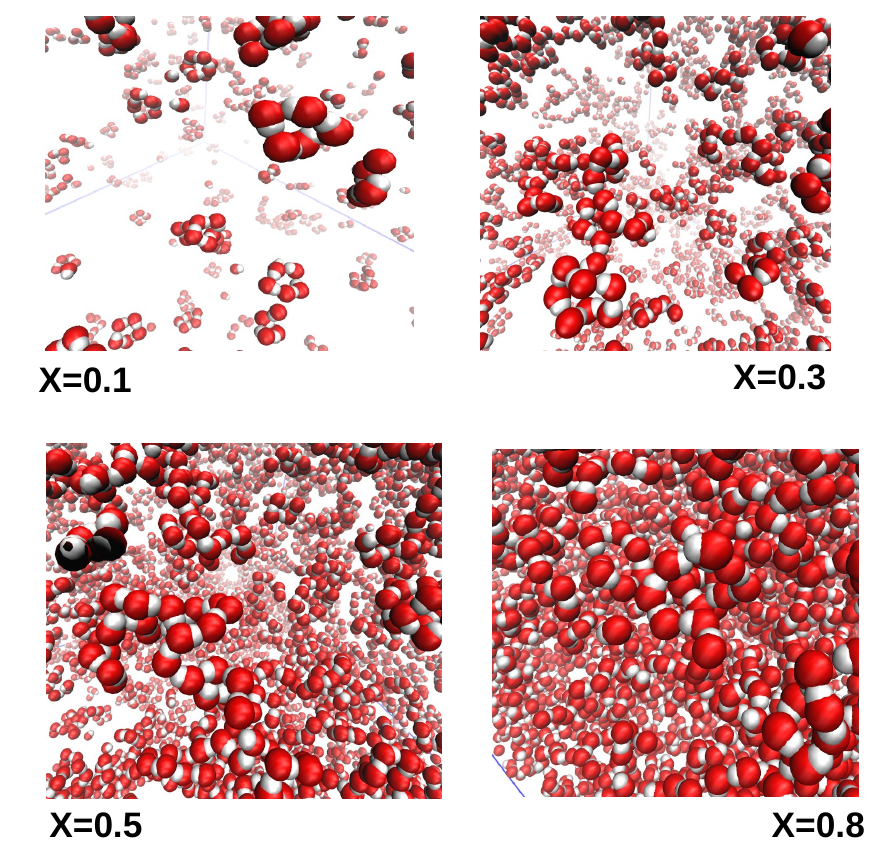}
\par\end{centering}
\caption{Zoom over the ethanol hydroxyl group clusters for the cases shown
in Fig.4. (Box edges can be seen as blue lines).}
\end{figure}

Detailed studies of snapshots for the entire concentration range shows
that ethanol molecules start to spontaneously form these pentameric
micelles at the smallest ethanol concentrations studied herein, which
is $x=0.02$. These pentameric micelle structures persist until $x=0.3$,
after which larger aggregated structures take over. The visual inspection
of formed structure tends to confirm that there are 2 regime of clustering,
pentameric ethanol micelles for $x<0.3$ and larger clusters for higher
concentrations. We now confirm this through cluster and pair correlation
function studies.

\subsubsection{Cluster structure}

The cluster size probability distributions were calculated as:

\begin{equation}
P_{n}=\frac{\sum_{k}s\left(n,k\right)}{\sum_{k}\sum_{n}s\left(n,k\right)}
\end{equation}

where $P_{n}$ is the probability for the cluster formed of \emph{n}
sites, $s\left(n,k\right)$ represents the number of clusters of the
size $n$ in the configuration $k$. $P_{n}$ is obtained by averaging
the number $s\left(n,k\right)$ of clusters of size \emph{n} over
several such configurations. The cut-off distance ($r_{c}$) for a
calculation depends on the first minimum in the pair correlation function
of a particular site. For the oxygens in ethanol, the $r_{c}$ = 3.7
Å.

Fig.6 shows the size $s$ dependence of the cluster distributions
$P(s)$ in 2 different ways. Left panel (a) shows the distribution
as function of the cluster size, and for different ethanol concentrations.
What is apparent is the pentamer peak, which exists for all concentrations.
In addition, we observe the appearance of a shoulder-like feature
for large clusters, showing that the probability of larger clusters
is not negligible for ethanol concentrations above $x>0.3$. Moreover,
while the probability of monomers is higher than that of pentamers
for $x<0.3$, and becomes comparable or lower for $x>0.3$.

\begin{figure}[H]
\begin{centering}
\includegraphics[scale=0.3]{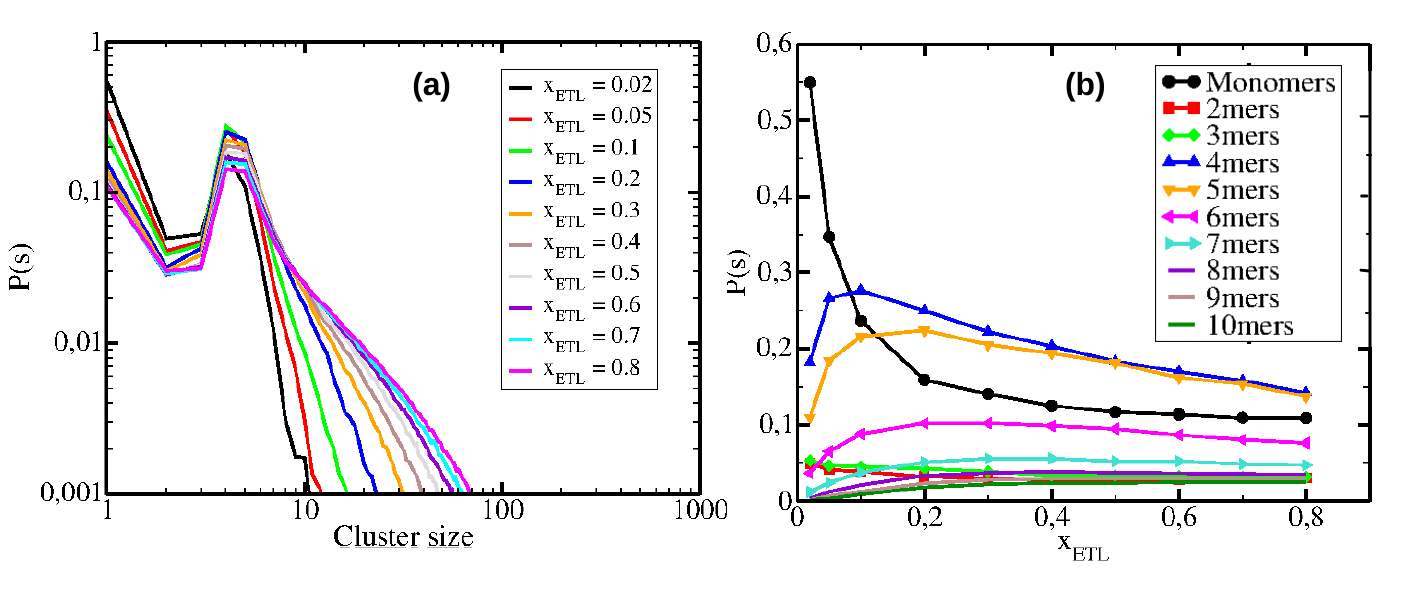}
\par\end{centering}
\caption{Cluster distribution probability function $P(s)$ as function of \textbf{(a)}
cluster size $s$ and \textbf{(b)} ethanol concentration. (See text
for details).}
\end{figure}

The right panel (b) shows $P(s)$ as function of ethanol concentration,
and for different types of $n$-mers. What is strikingly apparent
here is that, for the entire concentration range, there is a sharp
fall of P(s) from monomer to dimers and 3-mers, followed by a second
dramatic increase for $4$-mers and $5$-mers, followed by $6$-mers,
while higher $n$-mers are the same level as monomers and dimers.
An important feature is that the maximum of the curves seems to saturate
around $x=0.3$, indicating a turnover of cluster structures above
this concentration.

The cluster study complements the visual inspection of Fig.5 by showing
more details, such as for example the existence of 4-mers at small
concentrations, and not only pentamers, as the visual inspection may
suggest.

\subsubsection{Atom-atom pair correlation functions}

In order to further confirm the sharp separation of cluster structure
highlighted in the previous sections, we show in Fig.7 the evolution
of 3 typical pair correlation functions as a function of ethanol concentration.
The right panel shows the ethanol oxygen-oxygen correlation function
$g_{OO}(r)$ for all the ethanol concentrations calculated. Focusing
on the second peak, which represent second neighbour correlations,
hence the influence of clustering beyond the first neighbour peak,
we clearly a difference between two successive concentrations: the
gap is wider for $x<0.3$. In order to facilitate this observation,
the curve for x=0.3 is marked in thicker orange color.

\begin{figure}[H]
\begin{centering}
\includegraphics[scale=0.3]{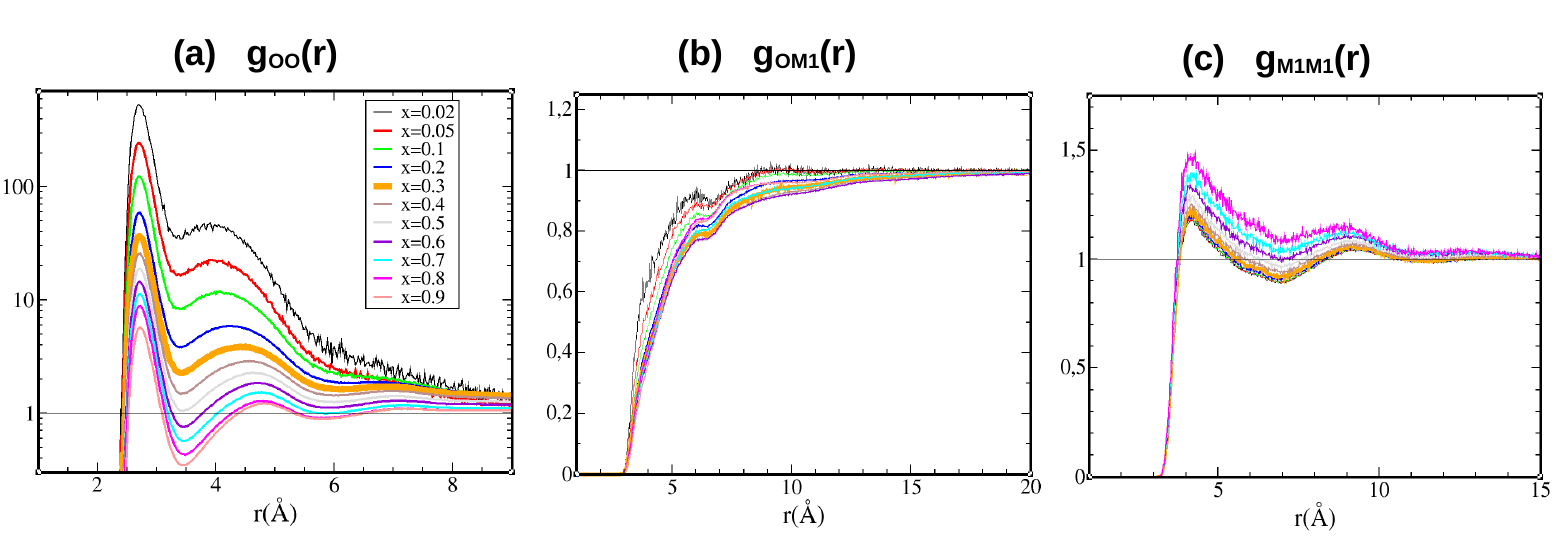}
\par\end{centering}
\caption{Ethanol concentration dependance of selected atom-atom pair correlation
functions. \textbf{(a)} Ethanol oxygen-oxygen pair correlation functions
$g_{OO}(r)$; \textbf{(b)} ethanol oxygen heptane 1st carbon group
cross correlations $g_{OM_{1}}(r)$: \textbf{(c)} heptane 1st carbon
group correlations $g_{M_{1}M_{1}}(r)$. The ethanol concentration
$x=0.3$ is marked at thicker orange line.}
\end{figure}

The narrower gap for $x>0.3$ can be interpreted as lesser difference
in the second neighbour correlations as a mark of insensitivity for
the concentration dependence. This is consistent with hydroxyl clusters
being of more varied shapes. In contrast, the almost similar gap for
x<0.3 shows a linear dependence in concentration, which would be expected
if clustering was the same and would only depend on the concentration
of the pentamers.

In order to further confirm this trend, we examine in the middle panel
of Fig.7 the cross species correlations $g_{OM_{1}}(r)$ between the
ethanol oxygen and the heptane first(last) carbon group termed M1.
These correlations are seen to be less concentration dependent than
the previously examined $g_{OO}(r)$ correlations. This is expected,
since heptane site are not charged, hence only Lennard-Jones like
correlations exist, which are less prominent then for those between
charged groups \cite{2015_PCCP_benzMH}. Nevertheless, we observe
that for $x<0.3$ all correlation functions are nearly superposed
(below the thick orange curve), and start to show appreciable differences
only when $x>0.3$. Again, this is fully consistent with the existence
of same type of clusters for x<0.3. Indeed, is clustered objects are
the same, the correlations between the carbon sites of heptane and
the meta objects would be nearly similar, and very weakly dependent
on x. In contrast, if the clustered objects are very different in
shape, and if this depends strongly on ethanol concentration, we would
indeed expect a larger concentration dependence of $g_{OM_{1}}(r)$.

The last right panel Fig.7 shows a weak concentration correlations
between the heptane last/first carbon site $g_{M_{1}M_{1}}(r)$, very
similar to that observed for $g_{OM_{1}}(r)$. First of al these are
Lennard-Jones like correlations, and correlations below x<0.3 are
also nearly superimposed, witnessing the same insensitivity to concentration
in presence of same type of aggregates of the ethanol.

If we reconsider now the experimental KBI in the light of this sharp
separation of the behaviour of the correlations for $x<0.3$ and for
$x>0.3$, we obtain the result shown in Fig.8, where the thick orange
vertical line at x=0.3 perfectly separates the

\begin{figure}
\begin{centering}
\includegraphics[scale=0.3]{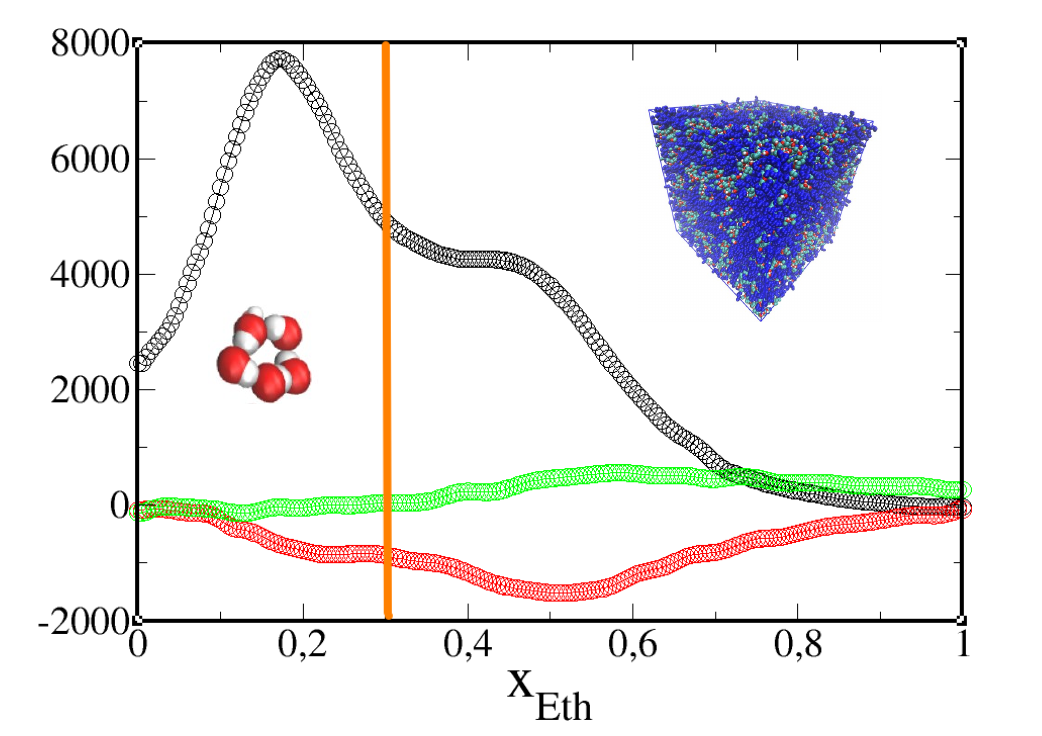}
\par\end{centering}
\caption{Illustration of the two ethanol concentration regions for each extremum
of the KBI, separated by $x=0.3$ (vertical orange line). Left part
for $x<0.3$ concerns ethanol micellar entities, while right part
$x>0.3$ concerns standard micro-segregated ethanol-heptane domains.}
\end{figure}

concentrations under the first KBI extremum for x$<0.3$, where mostly
pentameric ethanol micelles are observed, from the concentrations
under the second weaker peak at right for $x<0.3$, for which usual
micro-segregation of the 2 species is observed. In order to fully
confirm this picture of sharp separation between a meta-object mixture
and a usual mixture, we need to calculate the KBI from the simulations
and reproduce the same KBI shapes as observed in experiments.

\subsection{Kirkwood-Buff integrals from simulations}

The evaluation of the KBI through computer simulation requires that
the asymptotes of the various site-site correlations are well defined
and converging to 1 as expected. Recent investigations have shown
that there are 2 major obstacles. The first obstacle is that computer
simulations conducted in the Canonical or Isobaric ensemble cannot
lead to the proper asymptote 1, this value being reached only in the
Grand Canonical ensemble \cite{KBI_LP_correction,Salacuse,Salacuse2}.
The second obstacle concerns the existence of micro-segregation, whose
spatial extent and slow kinetics alters the statistics of the correlations
at large atom-atom separations. Both cases are illustrated below,
where we examine both the tail of the correlation functions $g_{ab}(r),$but
also the so-called running KBI defined as \cite{KBI_Kirkwood_Buff_seminal}:
\begin{equation}
G_{ab}(r)=4\pi\int_{0}^{r}ds\,s^{2}\left[g_{ab}(s)-1\right]\label{RKBI}
\end{equation}
which, when $r$ is large enough, is expected to converge to the KBI
defined in Eq.(\ref{KBI}).

In a first example, we examine the case of ethanol concentration $x=0.7$,
for which we have shown above that micro-segregation is dominant.
The typical order parameter for micro-segregation is $g_{OM_{1}}(r)$,
the cross species pair correlation between the ethanol oxygen atom
and the first/last carbon group atom of heptanol. Fig.9 shown this
function in the main panel for 3 different runs of 5ns each. The strong
depletion between adverse ethanol and heptane nano-domains is clearly
visible, as all 3 curves stay below 1 until $20\mathring{A}$ or so.

\begin{figure}[H]
\begin{centering}
\includegraphics[scale=0.3]{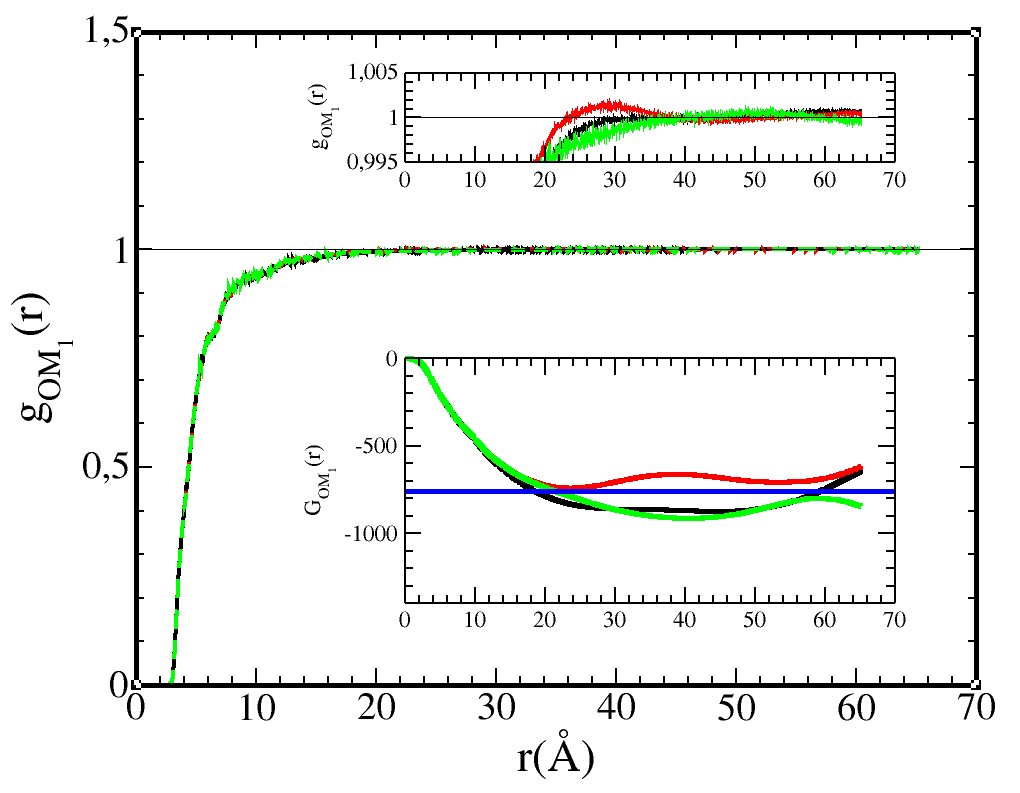}
\par\end{centering}
\caption{Long range tail contribution for ethanol-heptane cross correlation
function $g_{OM_{1}}(r)$ for ethanol concentration $x=0.7$ shown
in the main panel for 3 different runs of 5ns each. The upper inset
is a close-up of the tail part. The lower inset shows the respective
RKBI $G_{OM_{1}}(r)$ functions, along with the experimental KBI (in
blue).}
\end{figure}

On this scale, all 3 curves seem nearly similar, which is what is
expected from statistics. However, the zoom on tail region in the
upper inset shows that there are visible differences between the 3
runs, as the segregation of domain is not the same in each of the
3 runs. What is observed is the typical domains oscillatory correlations,
with half-period about $20\mathring{A}$, which is the depletion extent.
All 3 curves oscillated about the asymptote 1, hence the first obstacle
mentioned is not really apparent. The lower inset shows the corresponding
running KBI $G_{OM_{1}}(r)$, as well as the expected experimental
value. It is seen that the various runs have an oscillatory asymptotic
feature, witnessing the domains alternation, but oscillate around
the experimental value. This result shows the rather excellent agreement
between the experimental KBI and the calculated one.

In the second example, we examine the case of ethanol concentration
$x=0.2$, for which we expect an homogeneous distribution of ethanol
micelles in the midst of heptane solvent. In this case, micro-segregation
is very different in that it generates these micelles. Fig.10 shows
the same order parameter $g_{OM_{1}}(r)$ in the main panel, and for
5 different runs of 5ns each.

\begin{figure}
\begin{centering}
\includegraphics[scale=0.3]{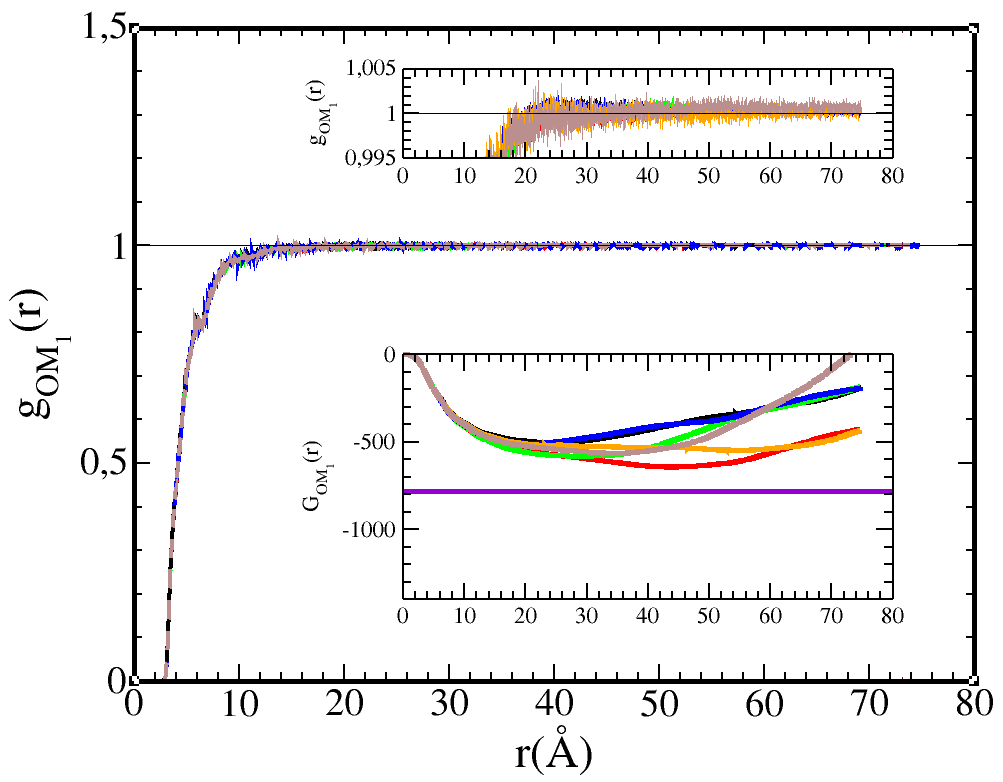}
\par\end{centering}
\caption{Long range tail contribution for ethanol-heptane cross correlation
function $g_{OM_{1}}(r)$ for ethanol concentration $x=0.2$ shown
in the main panel for 5 different runs of 5ns each. The upper inset
is a close-up of the tail part. The lower inset shows the respective
RKBI $G_{OM_{1}}(r)$ functions, along with the experimental KBI (in
purple).}
\end{figure}

This function again shows appreciable depletion, but of the size of
the ethanol micelles, which is more about $12\mathring{A}$. But now,
the zoom of the tail in the upper inset shows that there are no domain
oscillations, which is indeed expected if the ethanol droplets are
homogeneous distributed. However, we observe that all the asymptote
are clearly shifted upwards from 1. This is a direct manifestation
of the first obstacle mentioned above, which occurs here since our
calculations are in the isobaric ensemble instead of the Grand canonical
ensemble. It can be shown that the asymptote of the correlation functions
has the following form 
\begin{equation}
\lim_{r\rightarrow\infty}g_{ab}(r)=1-\frac{\epsilon_{ab}}{N}\label{LPshift}
\end{equation}
where $N$ is the number of particles in the simulation box, and the
value of $\epsilon_{ab}$ depend both on species $a$,$b$ and the
statistical ensemble \cite{KBI_LP_correction,Salacuse,Salacuse2}.
For like correlations the shift is downwards from 1, and for unlike
it is upwards \cite{2015_PCCP_benzMH}. The RKBI are shown in the
lower inset, and are seen to lie above the experimental value, and
not to have the expected flat asymptote for a proper defintition of
the KBI value for $G_{OM_{1}}$. In order to correct for this ``spurious''
behaviour, we simply multiply the pair correlation function with the
appropriate coefficient $\gamma_{ab}=1/\left(1-\frac{\epsilon_{ab}}{N}\right)$
before applying Eq.(\ref{LPshift}) to $\bar{g}_{OM_{1}}(r)=\gamma_{OM_{1}}g_{OM_{1}}(r)$,
which now has the correct asymptote 1. The corresponding results are
shown in Fig.11.

\begin{figure}[H]
\begin{centering}
\includegraphics[scale=0.3]{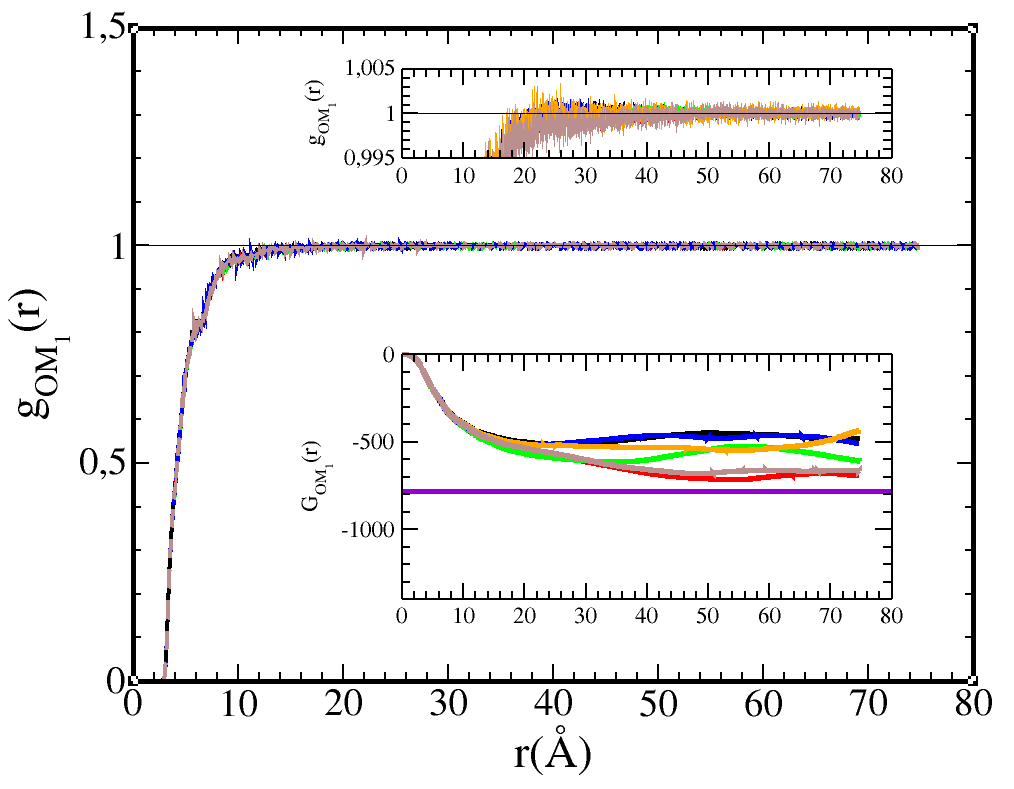}
\par\end{centering}
\caption{Tail corrected version of pair correlation functions shown Fig.10
(see text)}
\end{figure}

Due to the extreme smallness of $\frac{\epsilon_{ab}}{N}$, $\gamma_{ab}$
is nearly 1 and the main panel shows no difference with respect for
Fig.10. However, while the upper inset confirms that the correct asymptote
1 has been set by the operation above, the lower inset shows that
all the RKBI have now flat asymptotes, even though it is not quite
the value set by the experiments. This discrepancy is expected, since
the ethanol aggregation in the real mixtures may not be well represented
by the model simulations.

For the calculations of the KBI, from the cross KBI $G_{OM_{1}}$
we extract D from Eq.(\ref{Dab}) and calculate the like KBI $G_{OO}$
and $G_{M_{1}M_{1}}$ using Eq.(\ref{Daa}). Indeed, the simulations
results for these KBI do not often converge to the consistency for
D in Eqs.(\ref{Daa},\ref{Dab})), and the above trick is one way
to get around this problem. The origin of the inconsistency is due
to system size problems, since the like domain statistics require
larger simulations, with N=32000 particles or more \cite{SIM_Patey_Gupta_TBA,SIM_Patey_Overduin_TBA,AUP_Patey_TBA}.
When these operations are conducted for all ethanol concentrations,
we obtain the result shown in Fig.12, which shows a remarkable agreement
with the experimental results.

\begin{figure}
\begin{centering}
\includegraphics[scale=0.3]{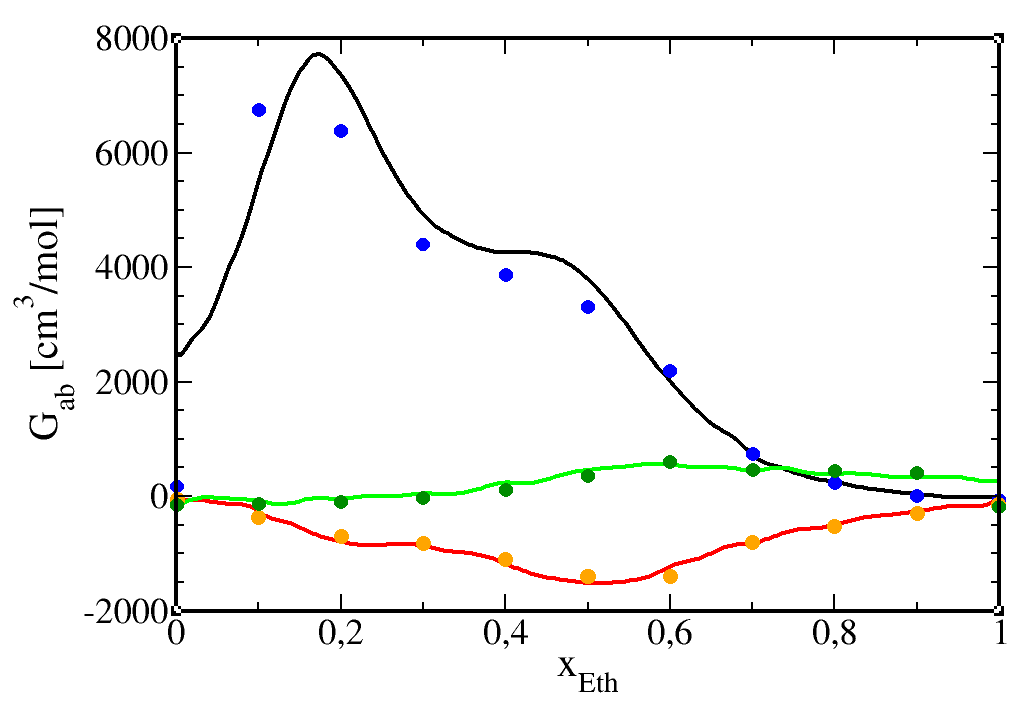}
\par\end{centering}
\caption{Comparison between the calculated KBI from simulations versus the
experimental ones \cite{KBI_Zielkiewicz}.}
\end{figure}

The very low ethanol concentrations for $x=0.02$ and $x=0.05$ pose
particular problems for the proper evaluation of the asymptotes. Indeed,
the corresponding RKBI have very distorted shapes, which cannot be
arranged to look flat by simple shifting of the asymptotes of the
pair correlation functions. These distortions, which vary considerably
from one run the other, witness the strong kinetics of the ethanol
micelle formation. Since these micelles are about $15\mathring{A}$
in diameter, while the half box length extends to $70\mathring{A}$,
if the micelles were replaced by soft spheres of same diameter, the
statistics on the tail of the soft sphere correlations would be much
better. Therefore, it is really the micelle-monomer exchange which
affect those statistics at very small ethanol concentrations.

\section{Discussion and Conclusion}

It is not a priori obvious that the segregation of ethanol in small
concentration in oil-like solvent should evolve in a discontinuous
manner into the micro-segregation as this concentration is increased,
leading to the separation illustrated in Fig.8. For instance, in a
previous study of ethanol-benzene mixtures \cite{2015_PCCP_benzMH},
we have found a single extremum shaped KBIs. Yet, this study has shown
that small micelle-like aggregates at low ethanol mole fraction mixtures.
This difference suggests that closed carbon group molecules, such
as benzene, affect differently the cluster structural changes from
micelle to domain, than chain-like shaped alkanes. Indeed, these latter
molecules can more easily merge with the micellar alkane corona than
the disc-shaped benzene molecules. In order to explain this apparent
discrepancy, we hypothetize here that it is solvent shape induced
depletion entropic effects \cite{DepltAsakura1954,DepletSapir2014,DepletSapir2015}
which affect the structural transition between the ethanol micelles
regime and the micro-segregated domain regime. Following this hypothesis,
chain-like solvent would allow a smoother transition between micelles
and domains, hence clearly separating the two manifestations in terms
of fluctuations, thus leading to separate fluctuation regimes.

In the abstract, we mention that water tert-butanol mixtures equally
show two extrema in KBI. This was indeed reported our previous work\cite{AUP_alcohol_KBI}
, where the KBI calculated by us (see Fig.6b in Ref.\cite{AUP_alcohol_KBI})
were obtained by Y. Koga from the vapour pressure measurements of
the chemical potentials (see the plot of  $D$ in Fig.6a of Ref.\cite{AUP_alcohol_KBI})
. The double extrema is absent from the KBI data from other authors
(see Fig.6b in Ref.\cite{AUP_alcohol_KBI}), but present in the SANS
data obtained in the same work. Interestingly, in boths cases, the
double extrema occurs clearly only for the water-water KBI. In addition,
the equivalent of the micelle extrema - in the present case it would
be tbutanol micellar aggregates, occurs for very small tbutanol concentrations
in the interval $0.1<x<0.2$, which is similar to that observed in
the present work for an entirely different system. Interestingly,
the aggregation in aqueous tbutanol is not clearly understood to date,
despite several experimental \cite{Winkler2019,Sedlak2014} and simulation
\cite{MH_Bagchi_TBA_Wat,SIM_Patey_Gupta_TBA,AUP_Patey_TBA,Cerar2020}
investigations.

From these two points, it appears that the topic of the existence
of camel back shaped KBI deserves further investigations as to the
conditions where it might occur.

To conclude, the present study illustrates the difference between
concentration fluctuations and micro-segregation both in the microscopic
and macroscopic thermodynamic level. It is generally believed that
micro-segregation would correspond to a more microscopic $k\neq0$
part of the concentration fluctuations, related to the finite extent
of segregated domains, while concentration fluctuations would be their
thermodynamic limit, which correspond to $k=0$. The present study
shown that the KBI, which are related to the macroscopic $k=0$ limit
of the structure factors, themselves contain the difference in both
manifestations. This was illustrated in Fig.8, through the sharp separation
between concentration fluctuations of meta-objects (the ethanol micelles)
for $x<0.3$ and the micro-segregation regime for $x>0.3$. To be
more specific, while micro-segregation concerns the initial microscopic
objects, namely ethanol and heptane molecules, the concentration fluctuations
described here concern the meta-objects formed by the ethanol micelles,
themselves floating in molecular heptane solvent. This study has been
possible, precisely because heptane is an inert solvent, due to its
uncharged carbon group sites, which cannot form associated entities.
In this context, heptane concentration fluctuations are very small,
and the neutrality of this oil-solvent enhances the charge association
of ethanol molecules. This is not possible to observe with water-solvent
and small alcohol molecules, because both species tend to associate.
In order to observe a similar phenomena in water, one requires much
larger solutes molecules, such as surfactants, which can form micelles
and other self-assembled objects.

The present study also illustrates the importance of the pair correlation
functions and their asymptotes, and through these quantities, the
issues related to domain segregation and concentration fluctuations
in finite size simulations, in particular through the Lebowitz-Percus
shift of the asymptotes, illustrate here for the fluctuations between
meta-objects, as opposed to the usual illustration for molecular objects.

\section*{Appendix}

Eq.(2) in the Kirkwood-Buff paper \cite{KBI_Kirkwood_Buff_seminal}
expresses the derivative of the chemical potential $\mu_{i}$of species
i with respect to its mole fraction $x_{i}$ in terms of the KBI,
which we rewrite below using a trivial rearrangement of the original
equation as: 
\[
\left(\frac{\partial\beta\mu_{i}}{\partial x_{i}}\right)_{T,P}=\frac{1}{x_{i}\left[1+\rho_{j}x_{i}\left(G_{11}+G_{22}-2G_{12}\right)\right]}
\]
The Ornstein-Zernike equation for a binary mixture can be written
in a matrix equation as 
\[
\left(\begin{array}{cc}
S_{11} & S_{12}\\
S_{12} & S_{22}
\end{array}\right)\left(\begin{array}{cc}
C_{11} & C_{12}\\
C_{12} & C_{22}
\end{array}\right)=I
\]
where $I$ is the identity matrix, and where the structure matrix
elements are $S_{ij}(k)=\delta_{ij}+\sqrt{\rho_{i}\rho_{j}}\tilde{h}_{ij}(k)$
and the C-matrix elements are defined in terms of the direct correlation
functions as $C_{ij}(k)=\delta_{ij}-\sqrt{\rho_{i}\rho_{j}}\tilde{c}_{ij}(k)$,
with the Fourier transform notation and definition $\tilde{f}(k)=\int d\vec{r}f(r)\exp(i\vec{r}.\vec{k})$.
Since the KBI are defined as in Eq.(\ref{KBI}), one has $G_{ij}=\tilde{h}_{ij}(k=0)$,
and $S_{ij}=\delta_{ij}+\sqrt{\rho_{i}\rho_{j}}G_{ij}.$ Therefore,
one can rewrite the equation above as (using $\rho_{i}=x_{i}\rho)$
\[
\left(\frac{\partial\beta\mu_{i}}{\partial x_{i}}\right)_{T,P}=\frac{1}{x_{i}\sqrt{x_{1}x_{2}}\left[S_{11}+S_{22}-2S_{12}\right]}
\]
which can be rewritten by using the OZ equation as 
\[
\left(\frac{\partial\beta\mu_{i}}{\partial x_{i}}\right)_{T,P}=\frac{\det C}{x_{i}\sqrt{x_{1}x_{2}}\left[C_{11}+C_{22}+2C_{12}\right]}
\]
Using Eq.(\ref{D}) which defines the KBI coefficient $D$, one has
the relation relating D to the determinant of C, which controls the
stability limit of a mixture \cite{Textbook_Hansen_McDonald}, as
written in Eq.(\ref{D-OZ}) 
\[
D=\alpha\left[C_{11}C_{22}-C_{12}^{2}\right]
\]
with $\alpha=1/(\sqrt{x_{1}x_{2}}\left[C_{11}+C_{22}+2C_{12}\right]).$

\bibliographystyle{unsrt}
\bibliography{camel_kbi}

\end{document}